\documentclass[twoside,11pt]{article}

\usepackage{cjp}
\usepackage{graphics}
\begin{document}

%\mainmatter              % start of the contributions

\title{Experimental Investigation of the Competing Orders and Quantum\\
Criticality in Hole- and Electron-Doped Cuprate Superconductors}

\author{N.-C. Yeh$^1$, C.-T. Chen$^1$, V. S. Zapf$^1$, A. D. Beyer$^1$, C. R. Hughes$^1$,\\
M.-S. PARK$^2$, K.-H. KIM$^2$, and S.-I. LEE$^2$}
\address{$^1$Department of Physics, California Institute of Technology, Pasadena, CA 91125, USA\\
$^2$Pohang University of Science and Technology, Pohang 790-784, Republic of Korea}

%\date{July 5, 2004}
\maketitle              % typeset the title of the contribution

\begin{abstract}
We investigate the issues of competing orders and quantum criticality
in cuprate superconductors via experimental studies of the 
high-field thermodynamic phase diagrams and the quasiparticle
tunneling spectroscopy. Our results suggest substantial field-induced 
quantum fluctuations in all cuprates investigated, 
and their correlation with quasiparticle spectra 
implies that both electron- (n-type) and 
hole-doped (p-type) cuprate superconductors are in close proximity 
to a quantum critical point that separates a pure superconducting (SC) 
phase from a phase consisting of coexisting SC and a competing order. 
We further suggests that the relevant competing order is likely 
a spin-density wave (SDW) or a charge density wave (CDW), 
which can couple to an in-plane Cu-O bond stretching 
longitudinal optical (LO) phonon mode in the p-type cuprates 
but not in the n-type cuprates. This cooperative interaction may 
account for the pseudogap phenomenon above $T_c$ only in the p-type
cuprate superconductors.  
\end{abstract}


\section{Introduction}
The pairing mechanism of high-temperature superconducting cuprates  
remains elusive to date, largely because of the existence of competing 
orders in the ground state of these strongly correlated electronic 
systems~\cite{Zhang97,Sachdev03,Kivelson03}, which gives rise to 
lack of universality in a variety of fundamental physical 
properties~\cite{Yeh02a,Yeh01,Chen02}. Amongst different scenarios 
for cuprate superconductivity, spin- and phonon-mediated pairing 
mechanisms have been most discussed~\cite{Yeh02a,Scalapino95}. 
The former is considered feasible because of the proximity 
of all cuprate superconductors to a Mott antiferromagnetic
(AFM) insulating state. The latter scenario receives attention
because of the correlation between significant softening 
of a longitudinal optical (LO) in-plane Cu-O stretching 
phonon mode and the occurrence of superconductivity 
in some of the p-type cuprates~\cite{Tachiki03}. Given that 
cuprate superconductors are strongly correlated electronic 
systems, the spin, charge and lattice degrees of freedom 
are likely all intertwined, and can cooperate at times
to yield enhanced collective phenomena. 

In this work, we attempt to address the issues of competing order
and quantum criticality via experimental studies of the high-field 
thermodynamic phase diagrams and quasiparticle spectroscopy of 
various p- and n-type cuprates. Our findings of strong 
field-induced quantum fluctuations in all cuprates 
and the corresponding correlation with quasiparticle spectroscopy 
suggest that cuprate superconductors, regardless of being 
electron- or hole-doped, are all in close proximity to 
a quantum critical point (QCP)~\cite{Sachdev03,Vojta00} 
that separates a pure superconducting (SC) state from a coexisting state 
of SC and a competing order. We shall discuss the relevant competing orders
in the context of existing experimental evidence and theory, and also examine 
the LO phonon-mediated pairing scenario. 

\begin{figure}[!t]
\centerline{\includegraphics{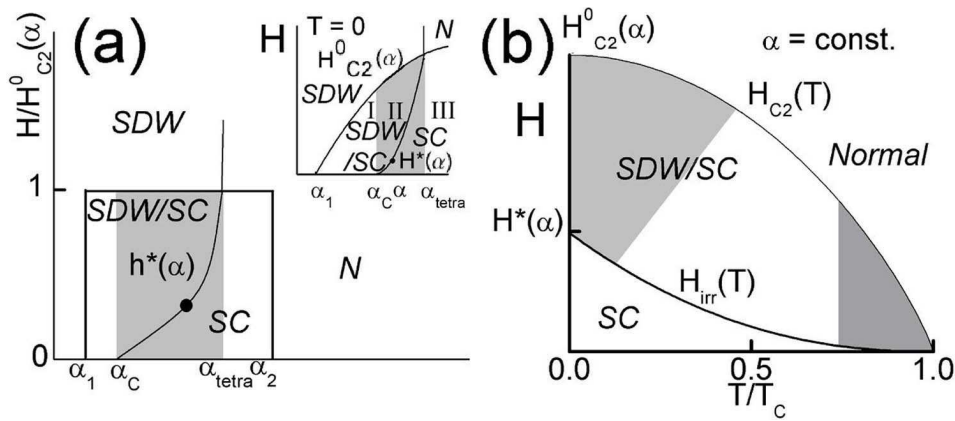}}
\figcaption{Schematic phase diagrams of cuprate superconductors with 
competing SDW and SC: (a) Reduced
field $h = (H/H_{c2}^0)$ vs. material parameter ($\alpha$) phase
diagram at $T = 0$ (main panel), transformed from the phase 
diagram in the inset\cite{Demler01}. Here $h^{\ast} (\alpha) \equiv
H^{\ast}(\alpha)/H_{c2}^0$ denotes the phase boundary that
separates a pure SC from a coexisting SDW/SC phase, and $N$ 
denotes the normal state. (b) $H(\alpha)$ vs. $(T/T_c)$ 
phase diagram for a given $\alpha$ value. The shaded
regime between $H_{irr}(T \to 0)$ and $H_{c2}(T \to 0)$ is 
associated with strong quantum fluctuations, and that between 
$H_{irr}(T \to T_c)$ and $H_{c2}(T \to T_c)$ is associated
with strong thermal fluctuations.}
\end{figure}

\section{Competing orders and quantum criticality}
Cuprate superconductors are doped Mott insulators with strong electronic 
correlation that can result in a variety of competing
orders in the ground state~\cite{Zhang97,Sachdev03,Kivelson03,Yeh02a}. 
In contrast to conventional superconductivity where the relevant low-energy
excitations are dominated by quasiparticles, the presence of competing 
orders in the cuprates implies that additional channels of low-energy excitations,
such as collective modes in the form of SDW~\cite{Demler01,ChenY02}, 
CDW~\cite{Kivelson03}, or staggered flux (also known as the d-density 
wave DDW)\cite{Kishine01}, can either coexist with SC 
or be induced by increasing magnetic field~\cite{Demler01,ChenY02}, 
temperature~\cite{ChenHY04}, disorder~\cite{Sachdev03,Polkovnikov02,Chen03}, 
currents~\cite{Yeh04a,Yeh04b}, etc. We therefore expect the
cuprate superconductors to exhibit significant quantum fluctuations
and reduced SC stiffness because of the existence of multiple channels 
of low-energy excitations. Indeed, macroscopically 
all cuprate superconductors are known to be extreme type-II 
superconductors~\cite{Blatter94}, a reflection of the weak SC stiffness 
associated with these doped Mott insulators. In addition, the 
presence of competing orders are likely responsible for the lack 
of universality in the pairing symmetry, pseudogap, quasiparticle 
spectral homogeneity, and commensuration of the low-energy spin 
excitations~\cite{Yeh02a,Yeh01,Chen02}. The proximity 
of cuprate SC to competing orders also result in quantum 
criticality~\cite{Zhang97,Sachdev03,Kivelson03,Yeh02a,Vojta00}.
Hence, it is instructive to investigate the degree of quantum 
fluctuations and the corresponding quasiparticle spectroscopy 
of cuprate superconductors in order to unravel the dominant 
competing order in the ground state. 
 
As a heuristic example, we consider a scenario that cuprate 
superconductivity occurs near a non-universal QCP at 
$\alpha = \alpha _c$ that separates the pure SC phase from a phase 
with coexisting SC and SDW~\cite{Demler01,ChenY02}, 
where $\alpha$ is a material-dependent parameter that may 
represent the doping level~\cite{Sachdev03,Vojta00}, the electronic anisotropy, 
spin correlation, orbital ordering, the on-site Coulomb repulsion~\cite{ChenHY04}, 
or the degree of disorder for a given family of cuprates~\cite{Sachdev03,Vojta00},
as illustrated in the inset of Fig.~1(a). The scenario of SDW 
as the relevant competing order can be rationalized 
by the proximity of cuprate SC to the Mott AFM, and also 
by experimental evidence for spin fluctuations in the SC state of  
cuprates~\cite{Wells97,Lake01,Mook02,Yamada03}. We stress here, however, 
that our experimental investigation and analysis given below can be 
generalized to other competing orders without restricting to the SDW.

The $T = 0$ phase diagram in Fig.~1(a) shows that in the absence of 
magnetic field $H$, the ground state consists of a pure SDW 
phase if $\alpha < \alpha _1$, a SDW/SC coexisting state if 
$\alpha _1 < \alpha < \alpha _c$ (Regime I), and a pure SC phase if 
$\alpha _c < \alpha \le \alpha _{tetra}$ (Regime II) or 
$\alpha _{tetra} < \alpha < \alpha _2$ (Regime III), where
$\alpha _{tetra}$ denotes a tetra-critical point. 
Upon applying magnetic field, spin fluctuations can be induced due 
to magnetic scattering from the $S = 1$ excitons centered around 
the vortex cores~\cite{Demler01}, which can be delocalized with 
increasing field and lead to a stabilized SDW coexisting with SC for 
$H^{\ast} (\alpha) < H < H_{c2}^0 (\alpha)$ in Regime II, 
where $H_{c2}^0(\alpha)$ denotes the upper critical field of a given
sample at $T = 0$. To quantify the degree of field-induced quantum 
fluctuations, we consider the normalized quantity 
$h^{\ast} (\alpha) \equiv \lbrack H^{\ast}(\alpha)/H_{c2}^0 (\alpha) \rbrack$, 
which transforms the phase diagram in the inset of Fig.~1(a) to 
that in the main panel. Here the characteristic field 
$H^{\ast}(\alpha)$ for a given sample can be determined from 
thermodynamic measurements of the irreversibility line $H_{irr}(T,\alpha)$ 
at $T \to 0$, as illustrated in Fig.~1(b). The transformed phase
diagram in Fig.~1(a) indicates that a smaller magnitude of 
$h^{\ast}(\alpha)$ corresponds to a closer proximity to 
$\alpha _c$ for a cuprate in Regime II. On the other 
hand, for a cuprate in Regime I, SDW coexists with 
SC even in the absence of external fields, implying gapless SDW excitations 
(i.e. $\Delta _{SDW} = 0$) and strong excess fluctuations in the 
SC state. Thus, the magnitude of $h^{\ast}(\alpha)$ for 
a given cuprate correlates with its SC stiffness and anti-correlates 
with its susceptibility to low-energy excitations. Such correlations
can be independently verified via studies of the quasiparticle spectra 
taken with a low-temperature scanning tunneling microscope (STM).

\section{Experimental Approach and Results}
To investigate the conjecture outlined above, we employ in this work
measurements of the penetration depth $\lambda (T,H)$, magnetization
$M(T,H)$, and third-harmonic susceptibility $\chi _3 (T,H)$
on different cuprates to determine the irreversibility field
$H_{irr}(T)$ and the upper critical field $H_{c2}(T)$. The degree
of quantum fluctuations in each sample is estimated by
the ratio $h^{\ast} \equiv (H^{\ast}/H_{c2}^0)$, where $H^{\ast}$ 
is defined as $H^{\ast} \equiv H_{irr}(T \to 0)$. The $h^{\ast}$ 
values thus determined for different cuprates are compared with the
corresponding quasiparticle spectra taken with a low-temperature
scanning tunneling microscope (STM). 

The cuprates studied in this work include the n-type optimally 
doped infinite-layer cuprate $\rm La_{0.1}Sr_{0.9}CuO_2$ 
(La-112, $T_c = 43$ K)~\cite{Chen02} and one-layer $\rm Nd_{1.85}Ce_{0.15}CuO_{4-\delta}$ 
(NCCO, $T_c = 21$ K)~\cite{Yeh92}; the p-type optimally doped 
$\rm HgBa_2Ca_3Cu_4O_x$ (Hg-1234, $T_c = 125$ K) and 
$\rm YBa_2Cu_3O_{7-\delta}$ (Y-123, $T_c = 93$ K)~\cite{Yeh01,Yeh93}. 
These results are compared with data obtained by other groups on 
p-type underdoped Y-123 ($T_c = 87$ K)~\cite{O'Brien00},
over- and optimally doped $\rm Bi_2Sr_2CaCu_2O_{8+x}$
(Bi-2212, $T_c = 60$ K and 93 K)~\cite{Krusin-Elbaum04,Krasnov00};
and n-type optimally doped $\rm Pr_{1.85}Ce_{0.15}CuO_{4-\delta}$
(PCCO, $T_c = 21$ K)~\cite{Kleefisch01}. Details of the synthesis and
characterization of various samples have been given elsewhere for
Y-123~\cite{Yeh01,Yeh93}, NCCO~\cite{Yeh92}, La-112~\cite{Jung02a}, 
and Hg-1234~\cite{KimMS98,KimMS01}.

\subsection{Thermodynamic measurements of the high-field phase diagram}
To determine $H_{c2}(T)$, the penetration depths $\lambda (T,H)$ of La-112 
and Hg-1234 were measured in the pulsed-field facilities up to
65 Tesla at the National High Magnetic Field Laboratory (NHMFL) 
in Los Alamos by detecting the frequency shift $\Delta f$ 
of a tunnel diode oscillator (TDO) resonant tank circuit, 
with the sample contained in one of the component 
inductors~\cite{Zapf04}. Small changes in the resonant 
frequency can be related to changes in the penetration depth 
$\Delta \lambda$ by $\Delta \lambda \propto 
- \frac{\Delta f}{f_0}$~\cite{Zapf04}. 
In our case, the reference frequency is $f_0 \sim 60$ MHz and
$\Delta f \sim$ (0.16 MHz/$\mu$m)$\Delta \lambda$, as detailed 
in Ref.~\cite{Zapf04}. For polycrystalline samples at a constant
temperature $T$, the onset of deviation of the resonant frequency
from that of the normal state signals the maximum upper
critical field $H_{c2}^{ab}(T)$, as exemplified by
La-112. On the other hand, $H_{c2}^c(T)$ of a polycrystalline
sample can be determined by placing a grain-aligned sample in the
TDO resonant tank circuit, as exemplified in the inset of Fig.~2(a) 
for La-112. The $M(T,H)$ measurements were conducted 
in high magnetic fields (up to 50 Tesla in a $^3$He refrigerator) 
using a compensated coil in the pulsed-field facilities, and 
in lower DC fields using a Quantum Design SQUID magnetometer 
at Caltech. The irreversibility field $H_{irr}(T)$ was identified from the
onset of reversibility in the $M(T,H)$ loops, as exemplified in
the inset of Fig.~2(b) for $M$-vs.-$H$ data of La-112 and in the main panel 
for $M$-vs.-$T$ data of Hg-1234. The third-harmonic magnetic 
susceptibility $\chi _3 (T,H)$ measurements were also performed on Hg-1234
sample using a 9-Tesla DC magnet and Hall probe techniques~\cite{Reed95}.
The $\chi _3 (T,H)$ data measured the non-linear response of the
sample and were therefore sensitive to the occurrence of
phase transformation~\cite{Reed95}. These measurements were carried
out to independently verify the results from $M(T,H)$-vs.-$T$.

\begin{figure}[!t]
\centerline{\includegraphics{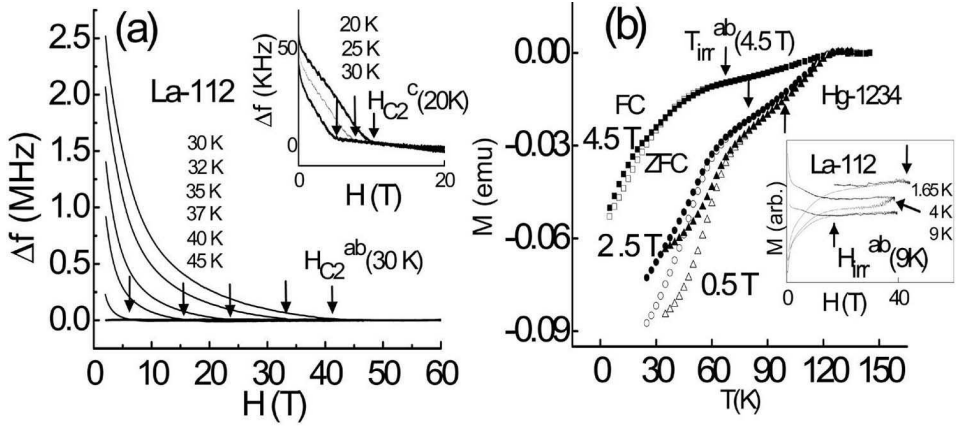}} 
\figcaption{(a) Selected data for changes in the resonant
frequency $\Delta f$ of the TDO tank circuit relative to the
normal state of a polycrystalline La-112 as a function of $H$ 
at various $T$. The estimated $H_{c2}^{ab}(T)$ for $H \parallel$ 
ab-plane are indicated by arrows. 
Similar measurements on a grain-aligned La-112 are shown in
the inset, which yield $H_{c2}^c(T)$. (b) Main panel:
representative zero-field-cooled (ZFC) and field-cooled (FC)
$M(T,H)$-vs.-$T$ data of Hg-1234 taken at $H = 0.5$, 2.5 and 4.5 Tesla,
with the corresponding $T_{irr}^{ab}(H)$ indicated by arrows, 
using the criterion 
$|M_{ZFC}(T_{irr},H) - M_{FC}(T_{irr},H)|/|M(T \to 0, H \to 0)|
\sim 1.5 \%$. The $T_{irr}^{ab}(H)$ values thus obtained 
are independently verified from measurements of $|\chi _3 (T,H)|$ 
(not shown, see Ref.~\cite{Yeh04b} for representative
data). Inset: $M(T,H)$-vs.-$H$ data on La-112 for 
$T = 1.65$, 4.0 and 9.0 K, where $H_{irr}^{ab} (T)$ are 
indicated by arrows~\cite{Zapf04}.}
\end{figure}

A collection of measured $H_{irr}^{ab}(T,\alpha)$ and $H_{c2}^{ab}(T,\alpha)$ 
curves normalized to the corresponding in-plane $H_{c2}^0 (\alpha) \sim 
H_p \equiv \Delta _{SC}^0/(\sqrt{2} \mu _B)$ for various 
cuprates are summarized in the main panel and the inset of Fig.~3(a),
respectively. Here $H_p$ is the paramagnetic field, and 
$\Delta _{SC}^0$ denotes the superconducting gap at $T = 0$, which can be
determined either directly from tunneling data or from the relation
$\Delta_{SC}^0 \sim 2.8 k_B T_c$ for under- and optimally doped p-type 
cuprates. The characteristic fields $(h^{\ast})$ 
of several cuprates are given in Fig.~3(b) for
comparison. We note that the Hg-1234 sample, while having
the highest $T_c$ and upper critical field (estimated at
$H_p \sim H_{c2}^{ab} \sim 500$ Tesla) among the cuprates
shown here, has the lowest reduced irreversibility line
$(H_{irr}^{ab}(T)/H_p)$. This phenomenon is not only due to the extreme 
two-dimensionality (2D) of Hg-1234~\cite{KimMS01} that leads 
to strong thermal fluctuations at high temperatures, but is also likely 
the result of its much closer proximity to the QCP, yielding strong 
field-induced quantum fluctuations at low temperatures. In the context of a 
$t$-$t^{\prime}$-$U$-$V$ model~\cite{ChenHY04}, we may consider the 
varying proximity of optimally doped cuprates Hg-1234, La-112, NCCO 
and Y-123 to the QCP as a manifestation of the varying 
on-site Coulomb repulsion $U$ and pair attraction $V$ of 
carriers in the CuO$_2$ plane; larger $U$ (smaller $V$) corresponds to 
closer proximity to a Mott AFM insulator. Thus, the quantity $(U/V)$ 
anti-correlates with $\alpha$. We also compare the field-induced 
anisotropy in the irreversibility fields ($H_{irr}^{ab}/H_{irr}^c$) 
among different cuprates, and find that all cuprates except Hg-1234 
exhibit increasing anisotropy as $T \to T_c^-$, as shown in the inset 
of Fig.~3(b). The implication of this anomalous temperature 
dependence requires further investigation.

\begin{figure}[!t]
\centerline{\includegraphics{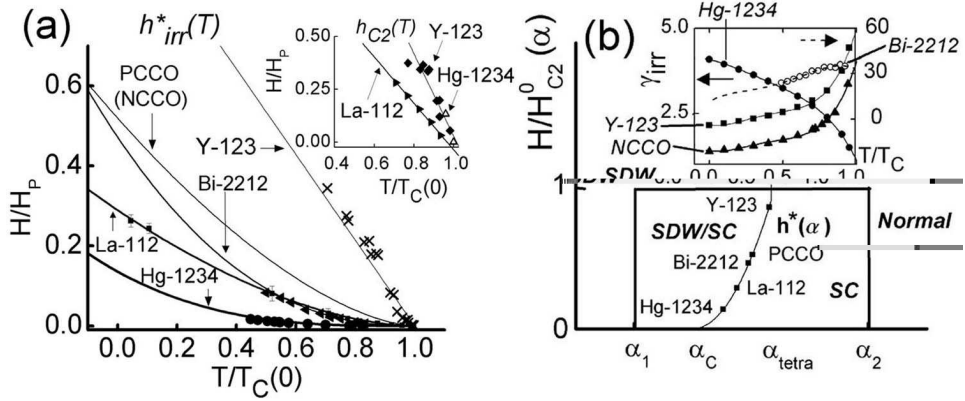}} 
\figcaption{(a) Comparison of the $h_{irr}(T)$ (main panel)
and $h_{c2}(T)$ (inset) vs. $(T/T_c)$ phase diagram for Hg-1234, 
La-112~\cite{Zapf04}, Bi-2212~\cite{Krusin-Elbaum04}, NCCO~\cite{Yeh92,Kleefisch01}, 
and Y-123~\cite{Yeh93,O'Brien00} with $H \parallel$ ab-plane, so that 
the corresponding $H_{c2}^0$ is limited by the paramagnetic field $H_p$. 
(b) Main panel: summary of the proximity of different cuprates
to a QCP at $\alpha _c$. The points along $h^{\ast}$ 
represent the reduced irreversibility fields obtained from data 
taken on Y-123, PCCO (NCCO), Bi-2212, La-112, and Hg-1234. Inset:
Temperature evolution of the anisotropy of the irreversibility
fields ($\gamma _{irr} \equiv H_{irr}^{ab}/H_{irr}^c$). The left (right) 
scale is for solid (open) symbols.} 
\end{figure}

\subsection{Quasiparticle spectroscopy}
To further verify the conjecture that cuprates with smaller
$h^{\ast}$ are in closer proximity to a QCP at $\alpha _c$ and
are therefore associated with a smaller SDW gap $\Delta _{SDW}$ and
stronger quantum fluctuations, we examine the
SC energy gap $\Delta _{SC} (T)$ and the quasiparticle
low-energy excitations of different cuprates. In Fig.~4(a),
we compare the $\Delta _{SC}(T)$ data of La-112, taken with a
low-temperature STM, with those of Bi-2212 and PCCO obtained 
from intrinsic tunnel junctions~\cite{Krasnov00}
and grain-boundary junctions~\cite{Kleefisch01}, respectively. 
We find that the rate of decrease in $\Delta _{SC}$ with 
$T$ follows the same trend as the decreasing $h^{\ast}$ values 
for PCCO ($\sim 0.53$), Bi-2212 ($\sim 0.45$), and La-112 ($\sim 0.24$).
These differences in $\Delta _{SC}(T)$ cannot be attributed
to different pairing symmetries, because the optimally doped 
La-112 and NCCO (PCCO) exhibit $s$-wave pairing symmetry
in the quasiparticle tunneling spectra~\cite{Chen02,Alff99}, and 
Bi-2212 is predominantly $d_{x^2-y^2}$-wave 
pairing~\cite{Renner98,Pan00,Hudson01}. Therefore, the sharp 
contrast in the $\Delta _{SC}(T)$ data between
La-112 and NCCO (PCCO) suggests that the proximity to the QCP
at $\alpha _c$ plays an important role in determining the low-energy
excitations of the cuprates. These experimental findings are 
consistent with recent $t$-$t^{\prime}$-$U$-$V$ model 
calculations~\cite{ChenHY04} for optimally doped p-type cuprates, 
which reveal that SDW can be induced thermally even if $H = 0$, 
similar to the field-induced SDW at $T = 0$, provided that the material 
parameter $\alpha$ falls within Regime II depicted in 
Fig.~1(a). An example of thermally induced SDW order parameter and 
the corresponding temperature dependence of the SC order parameter
is shown in the inset of Fig.~4(a), following Ref.~\cite{ChenHY04}.

\begin{figure}[!t]
\centerline{\includegraphics{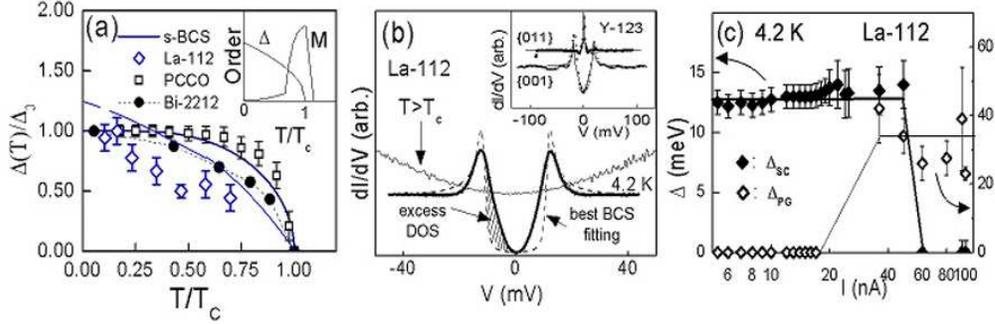}}
\figcaption{(a) Comparison of the temperature ($T$)
dependence of the normalized SC gap, $\lbrack
\Delta _{SC}(T) / \Delta _{SC}^0 \rbrack$, for NCCO
(PCCO)~\cite{Kleefisch01}, Bi-2212~\cite{Krasnov00}, and La-112. 
Inset: temperature evolution of the SC and SDW order parameters
($\Delta$ and $M$) derived from the $t$-$t^{\prime}$-$U$-$V$ model 
for an optimally doped cuprate with an effective $\alpha$ in 
Regime II of Fig.~1(a), after Ref.~\cite{ChenHY04}. 
(b) Comparison of the normalized quasiparticle spectrum of
La-112, obtained using a STM at $T = 4.2$ K, with the
best BCS fitting (dashed line), showing excess sub-gap and reduced
post-gap spectral weight. Also shown is the La-112 spectrum taken at
$T >\sim T_c$, revealing the absence of any quasiparticle pseudogap. 
Inset: comparison of the normalized quasiparticle tunneling 
spectra (points) of Y-123 for quasiparticle momenta along c-axis
and the nodal direction, taken using a STM at 4.2 K, with the 
generalized BTK fitting~\cite{Yeh01,Wei98} for $d_{x^2-y^2}$-wave 
superconductors (solid line). (c) Evolution of the SC gap $\Delta_{SC}$ 
and pseudogap $\Delta_{PG}$ with the tunneling current $I$ in 
La-112 at $T = 4.2$ K~\cite{Yeh04b}.}
\end{figure}

Further experimental confirmation for the correlation between $h^{\ast}$ 
and quantum fluctuations can be realized in the quasiparticle tunneling 
spectra (the differential conductance $dI/dV$ vs. biased voltage $V$). 
As exemplified in Fig.~4(b), we find that excess sub-gap spectral 
weight relative to the BCS prediction exists in La-112, although 
the quasiparticle spectra of La-112 are {\it momentum-independent}
and its response to quantum impurities is consistent with
$s$-wave pairing~\cite{Chen02,Yeh03a}. The phenomenon  
cannot be reconciled with simple quasiparticle excitations from 
a pure SC state, implying the presence of a competing order. 
In contrast, the quasiparticle spectra on optimally doped Y-123
(for quasiparticle energies $|E|$ up to $>\sim \Delta _{SC}$)
can be explained by the generalized BTK theory~\cite{Yeh01,Wei98}, as
shown in the inset of Fig.~4(b) for averaged quasiparticle tunneling
momenta along the c-axis and along the nodal direction of the
$d_{x^2-y^2}$-wave pairing potential. These results reveal 
that the spectral contribution from the competing order in 
Y-123 is insignificant up to $|E| \sim \Delta _{SC}$, in agreement 
with a much larger $h^{\ast}$ value and therefore much weaker
quantum fluctuations. Finally, by increasing the tunneling current
$I$, we observe suppression of the $\Delta _{SC}$ and emergence of
a pseudogap $\Delta _{PG}$ in the quasiparticle spectra of
La-112, as shown in Fig.~4(c), but not in Y-123, again confirming
the closer proximity of La-112 to a QCP. 

\section{Disorder effect, LO phonons, and pseudogap}

Next, we consider the effect of disorder on competing orders
and SC in the cuprates. Generally speaking, disorder reduces
the SC stiffness and shifts $\alpha$ from Regime II
toward Regime I~\cite{Vojta00}. In this context, a spatially 
inhomogeneous disorder potential can give rise to 
spatially varying $\alpha$ values in a cuprate. 
In particular, disorder can pin the fluctuating SDW 
more efficiently in 2D cuprates like Bi-2212 and Hg-1234
than in more 3D cuprates like Y-123 and La-112~\cite{Yeh03b},
so that regions with the disorder-pinned SDW can coexist with SC
in the former even if on average $\alpha > \alpha _c$. 
These randomly distributed regions of pinned SDW are 
scattering sites for quasiparticles at $T < T_c$ and for 
normal carriers at $T > T_c$. Our recent numerical 
calculations using the T-matrix approximation~\cite{Chen03} 
have shown that the quasiparticle interference spectra for
SC with coexisting disorder-pinned SDW differ fundamentally
from those due to pure SC with random point disorder. In 
particular, we find that the Fourier transformation (FT) of
the quasiparticle local density of states (LDOS) data, obtained 
from the STM studies of a slightly underdoped Bi-2212 at 
$T \ll T_c$~\cite{Hoffman02,McElroy03}, are consistent with
the superposition of spectra simulated for SC with random 
point defects and for SC with randomly pinned SDW regions.
In contrast, the FT-LDOS data  and the corresponding dispersion 
relations taken on a similar sample at 
$T > \sim T_c$~\cite{Vershinin04} are in good agreement
with our simulated results for scattering of holes from
disorder-pinned SDW. These findings suggest that a 
disorder-pinned SDW coexists with SC in Bi-2212 at low temperatures, 
and that only the SDW persists above $T_c$~\cite{Chen03}. 

The aforementioned numerical studies clearly demonstrate
the significant role of competing orders in determining
the physical properties of cuprate superconductors. In particular, 
disorder-pinned collective modes such as SDW can account 
for the strong spatial inhomogeneity observed in the quasiparticle 
spectra of 2D underdoped cuprates like Bi-2212 below 
$T_c$~\cite{Lang02}. Additionally, we suggest that 
collective modes such as SDW and CDW, which are associated with
spin/charge modulations along the Cu-O bonding directions, 
can couple well with the LO in-plane Cu-O stretching 
phonon mode in p-type cuprates at moderate temperatures, because 
the latter is known to involve substantial charge transfer from 
holes in the oxygen 2$p$-orbital to the copper 
3$d_{z^2}$-orbital, giving rise to a negative dielectric
constant and a local attractive potential~\cite{Tachiki03}. 
This cooperative effect can result in charge heterogeneity and 
stabilization of the collective SDW/CDW modes, yielding the 
pseudogap phenomenon~\cite{Timusk99} above $T_c$ in under- and 
optimally doped p-type cuprates. In contrast, we note that 
the LO in-plane Cu-O stretching phonon mode cannot incur 
significant charge transfer in the n-type cuprates as in the 
p-type, because electron doping in the former primarily resides 
on the copper site, giving rise to filled 3$d$- and 2$p$-orbitals 
in the Cu$^+$-O$^{2-}$ bonds, which do not favor phonon-induced 
charge transfer. We therefore argue that the quantum cooperation
effect between the LO phonons and the SDW/CDW collective modes is
insignificant in the n-type cuprates, which is consistent with the
absence of zero-field pseudogap in the quasiparticle tunneling
spectra of n-type cuprates, as exemplified in Fig.~4(b). 
More importantly, given that the LO phonon-induced attractive 
potential is not universal in the cuprates, and that similar 
phonon-induced charge transfer is present in other 
non-superconducting perovskite oxides~\cite{Tranquada02,Reichardt98},
the LO phonons are unlikely solely responsible for 
the superconducting pairing mechanism in the cuprates, even 
though they might enhance the $T_c$ values of the p-type cuprate
superconductors. 

\section{Summary and Remarks}

In summary, our experimental studies of the quasiparticle 
tunneling spectra and the thermodynamic high-field phase 
diagrams of p- and n-type cuprate superconductors reveal 
that significant magnetic field-induced quantum fluctuations 
exist in all cuprate superconductors, and that the degree of 
quantum fluctuations correlates with the magnitude 
of excess low-energy excitations as the result of competing 
orders in the ground state. These experimental results support 
the notion that the ground state of cuprates is in close proximity to 
a QCP that separates the SC phase from a coexisting SC and competing
order phase, and the competing order is likely associated with 
a form of collective modes such as the SDW or CDW. Our theoretical 
analysis further suggests that the presence of disorder-pinned SDW 
or CDW in highly 2D cuprates can account for the empirical LDOS both
below and above $T_c$. Additionally, the presence of SDW or CDW can 
couple with the LO phonons for the in-plane Cu-O stretching mode in p-type 
cuprates, because the latter can induce significant transfer of
holes from the oxygen 2$p$-orbital to the copper 3$d_{z^2}$-orbital,
giving rise to a local attractive potential. This quantum
cooperation can result in charge heterogeneity and is likely 
responsible for the pseudogap phenomenon in p-type cuprates. 
In contrast, the same LO phonons in n-type cuprates do not 
induce significant charge transfer, which may account for the 
long-range spectral homogeneity and the absence of zero-field 
pseudogap in the quasiparticle spectra of n-type cuprates. 
Hence, the conjecture of LO-phonon mediated pairing mechanism 
does not seem applicable to all cuprates. 

To further our understanding of the competing orders and quantum 
cooperation of collective modes, systematic studies of the 
correlation between the low-temperature high-field phase diagrams
and the low-energy excitations of more cuprates will be 
necessary. In particular, neutron scattering studies will 
be important for determining the SDW gaps and for establishing 
their relation to quantum fluctuations. 

%%%%%%%%%%%%%%%%%%%%%%%%%%%%%%%%%%%%%%%%%%%%%%%%%%%%%%%%%%%%%%%%%%%%%%%
\section*{Acknowledgment}
 
The work at Caltech is supported by the National Science Foundation
through Grants \#DMR-0405088 and \#DMR-0103045, and at the Pohang University
by the Ministry of Science and Technology of Korea.

% ---- Bibliography ----

\end{document}